# Construction of a Pragmatic Base Line for Journal Classifications and Maps Based on Aggregated Journal-Journal Citation Relations

*Journal of Informetrics* (in press)

Loet Leydesdorff,*[a] Lutz Bornmann,[b] and Ping Zhou[c]

**Abstract**
A number of journal classification systems have been developed in bibliometrics since the launch of the Citation Indices by the Institute of Scientific Information (ISI) in the 1960s. These systems are used to normalize citation counts with respect to field-specific citation patterns. The best known system is the so-called "Web-of-Science Subject Categories" (WCs). In other systems papers are classified by algorithmic solutions. Using the Journal Citation Reports 2014 of the Science Citation Index and the Social Science Citation Index (*n* of journals = 11,149), we examine options for developing a new system based on journal classifications into subject categories using aggregated journal-journal citation data. Combining routines in VOSviewer and Pajek, a tree-like classification is developed. At each level one can generate a map of science for all the journals subsumed under a category. Nine major fields are distinguished at the top level. Further decomposition of the social sciences is pursued for the sake of example with a focus on journals in information science (LIS) and science studies (STS). The new classification system improves on alternative options by avoiding the problem of randomness in each run that has made algorithmic solutions hitherto irreproducible. Limitations of the new system are discussed (e.g. the classification of multi-disciplinary journals). The system's usefulness for field-normalization in bibliometrics should be explored in future studies.

[a] * corresponding author; Amsterdam School of Communication Research (ASCoR), University of Amsterdam PO Box 15793, 1001 NG Amsterdam, The Netherlands; loet@leydesdorff.net
[b] Division for Science and Innovation Studies, Administrative Headquarters of the Max Planck Society, Hofgartenstr. 8, 80539 Munich, Germany; bornmann@gv.mpg.de
[c] Department of Information Resources Management, School of Public Affairs, Zhejiang University, No. 866 Yuhangtang Road, Hangzhou, 310058, China; pingzhou@zju.edu.cn

# 1. Introduction

If bibliometricians wish to normalize for differences in publication and citation behavior among fields of science, they use one field classification scheme or another. Since both WoS and Scopus are based on sets of journals, a classification of these journals provides an obvious candidate. For this purpose Thomson Reuters tags the journals with the "Web-of-Science Subject Categories" (WC), e.g. "chemistry, applied" or "biophysics." More than a single WC can be attributed to each journal in WoS.[4] An analogous journal classification system in terms of fields and subfields has been made available by Scopus (Wang & Waltman, 2016).[5] The use of these journal categories for normalization purposes has become accepted as "best practice" among bibliometricians (e.g., Rehn *et al*., 2014).

For example, InCites—a customized, web-based research evaluation tool developed by Thomson Reuters—routinely provides normalizations of citation impact using WCs for the delineation of the reference sets (e.g., Costas *et al*., 2010, at p. 1567). The Flemish ECOOM unit for evaluation in Leuven (SOOI), however, has developed another classification system for journals (Glänzel & Schubert, 2003). Other authors have refined the journal lists within specific WCs to enable a more precise evaluation of a given discipline (e.g., Van Leeuwen and Calero-Medina, 2012; cf. Bordons *et al*., 2004; Katz & Hicks, 1995).

---

[4] In the alternative classification developed since 1972 by Computer Horizon's Inc. for the *Science & Engineering Indicators* series of the NSF (Carpenter & Narin, 1973; Narin, 1976; Narin & Carpenter, 1972), a single category was attributed to each journal.
[5] The field/subfield classification of Scopus is available in the journal list from http://www.elsevier.com/online-tools/scopus/content-overview . WCs are available (under subscription) at http://images.webofknowledge.com/WOKRS56B5/help/WOS/hp_subject_category_terms_tasca.html .

Elsevier's *Scopus* introduced the SNIP indicator as an alternative to Thomson Reuters impact factor; SNIP is largely independent of structural assumptions about disciplines and specialties because the citing papers are used as the reference sets (Moed, 2010). Researchers at the Center for Science and Technology Studies in Leiden (CWTS) went one step further and proposed clustering the WoS at the level of documents as an alternative to journal classification and mapping (Waltman & van Eck, 2012). However, the 4000+ resulting clusters cannot easily be validated or reproduced (Klavans & Boyack, 2015; Leydesdorff & Bornmann, 2016).

Glänzel & Schubert (2003: 358) distinguish among (1) a cognitive approach when one classifies journal in terms of disciplines and specialties, (2) a pragmatic approach using journal classifications for the delineation of fields and subfields, and (3) a scientometric approach at the article level in which one tries to capture also the complexity of the system. This study can be considered as belonging to the second, that is, pragmatic approach. Using the Journal Citation Reports 2014 of the Science Citation Index and the Social Science Citation Index ($n$ of journals = 11,149), we examine options for developing a new system based on journal classifications into subject categories using aggregated journal-journal citation data. Ideally, a classification should be transparent and reasonably easy to reproduce outside the context of its production. As a second objective, a hierarchical classification can also be coupled to maps of the sciences at different levels of granularity (Zitt *et al*., 2005), so that one would be able to zoom in and out in order to distinguish among fields, sub-fields, sub-sub-fields, etc. Combining routines in VOSviewer and Pajek, a tree-like classification is developed in this study. At each level one can generate a map of science for all the journals subsumed under a category.

## 2. Algorithmic classifications

The further development of computer power and software makes it possible nowadays to generate algorithmically a comprehensive map and classification of the aggregated journal-journal relations provided by the Journal Citation Reports (JCR) of the (Social) Science Citation Index or similar data of Scopus (e.g., Gomez-Nuñez *et al*., 2014). Using 2012 data and two new algorithms (Newman & Girvan, 2004; Rosvall & Bergstrom, 2008), Rafols & Leydesdorff (2009) compared the resulting classifications with the WCs and with Schubert & Glänzel's (2003) revision as two content-based classifications. They found that the correspondences among the main categories are sometimes as low as 50% of the journals included; most of the mismatched journals appear to fall in areas in close proximity to the main categories. The results of the various decompositions are roughly consistent, but the overlap is imprecise (cf. Klavans & Boyack, 2009). The algorithmic constructs are more specific than the content-based classification of the ISI and SOOI, but the algorithms produce much more skewed distributions in terms of the number of journals per category.

In addition to the skew in the distributions generated in the algorithmic solutions—with potentially large tails of singletons—the randomness in each run makes the algorithmic classifications irreproducible from year to year (Lambiotte, *personal communications*, from 10 October 2008 to 16 December 2009). Consequently, it is unclear whether the differences in outcomes between two runs are due to relevant changes in the data or the randomness factor in the algorithm. This problem seemed unsolvable at the time. However, more recent developments in software development encourage us to make another attempt to construct the envisaged classification.

Among these new developments are:

1. The algorithms for the decomposition of large networks have been further developed since Newman & Girvan (2004). The programs of Blondel *et al*. (2008) and Waltman, van Eck, and Noyons (2010) for VOSviewer are seamlessly integrated in the context of Pajek, a program for the analysis and visualization of networks available in the public domain. These programs also provide modularity measures (*Q* and *VOS Quality*, respectively) as indicators of the decomposability of the data.

2. Pajek-files can function as a kind of currency for the transport of files among network programs such as Gephi, ORA, VOSviewer, UCInet, etc., each with their specific strengths. Moreover, in addition to its clustering and mapping algorithms, VOSviewer specifically allows for visualizing large networks, because the labels fade in and out with the level of granularity and without cluttering of the labels. The integration between Pajek and VOSviewer enables us to combine the options for network analysis, specific layouts (e.g., Kamada & Kawai, 1989), and statistics in Pajek (or UCInet) with the visualizations in VOSviewer.

3. Furthermore, the three-rings algorithm implemented in Pajek provides fast access to clique analysis (Batagelj & Zaveršnik, 2007; de Nooy & Leydesdorff, 2015). Cliques of three (or more) journals are the natural candidates for system formation through mechanisms of transitivity and triadic closure (Bianconi *et al*., 2014; Freeman, 1992 and 1996; Simmel, 1902);

When triads are considered as building blocks of systems, the clustering is agglomerative. In this study, we focus first on divisive clustering and postpone the analysis using triads to a next follow-up. Divisive clustering operates on the system and sorts similar elements together in subsystems, which can also be called partitions. Whereas the agglomerative clustering of triads ("cliques"; Hanneman & Riddle, 2005; cf. Freeman, 1996) can be otherwise parameter free using graph theory, at least two parameters need to be chosen in the case of divisive clustering: the clustering algorithm and a similarity criterion (e.g., the cosine values between each two patterns). The fast decomposition algorithms that we use in this study contain such parameters; both Pajek and VOSviewer allow for changing them. Since our purpose is not to search a parameter space for optimal configurations, but to develop a method to generate a classification system so that it can be produced, for example, for different years, we limit the analysis to default values in Pajek and VOSviewer.

Pajek provides a common framework for two decomposition algorithms denoted in this context as "VOS Clustering" (Van Eck *et al*., 2010) and the "Louvain Method" (Blondel *et al*., 2008), respectively. Both clustering routines begin with the choice of a random number. In VOSviewer itself the seed of the random number generator can be kept constant. As we shall see, this stabilizes the resulting number of clusters.

We will first compare the results of using either algorithm and explore the question of whether to use the largest component of the full data set or to use a threshold; for example, only citation relations that occur five or more times during a year. Alternating between Pajek and VOSviewer in iterations enables us in a second step to develop the envisaged dendrogram that can be mapped at different levels. We pursue the decomposition in greater detail for the cluster that contains *Journal of the Association for Information Science and Technology* (JASIST) and *Scientometrics*, an example chosen because we feel legitimated to validate results in this area.

A major disadvantage of a hierarchical classification is that each journal is classified in one of the categories (multiple assignments are not possible). In Section 7, we discuss this using (i) *PLoS ONE* as an example of a multidisciplinary journal and (ii) law journals which are attributed to two different branches of the dendrogram. Furthermore, we discuss the extension to the dynamic perspective.

**3. Data**

The two Journal Citation Reports (JCRs) 2014 contain 8,618 journals in the Science Citation Index and 3,143 in the Social Sciences Citation Index, respectively. However, the combined set covers 11,149 journals since 612 journals are included in both databases. We first generated the asymmetrical 1-mode matrix of these 11,149 journals cited (rows) versus citing (columns) from the database using dedicated routines. Of the 11,149 journals, 11,143 (> 99.9%) form a single largest component. The density of the network is 0.0217 or, in other words, 2.17% of the possible relations are realized, leading to 2,699,210 links. However, the average total degree is 484.207,[6] indicating that the network can not only be considered as a single (largest) component, but this component is also well-connected internally. The clustering coefficient of the network (CC1 in Pajek) is 0.220. This provides a measure for the transitivity in the network.

---

[6] The network is asymmetrical. As a graph, however, each outgoing line corresponds to an incoming one for another node. Thus, the average outdegree and indegree are both 242.103.

Of the approximately 2.7 million links only 112 are single citation relations. In the other cases, the database producer (Thomson Reuters) aggregates the long tails of the citation distribution with value one under the heading "All others." However, 55.5% of the links have a value of 2, 3, or 4. In a second matrix, these relatively weak citation relations (below five) were removed from the data. The largest component of this reduced network contains 11,087 vertices (99.4% of 11,149), but the number of links is now only 1,196,343 and the density is 0.010. The average degree is reduced to 215.810. However, only 62 journals are disconnected. In summary, this network is far more concentrated than the original one despite these minimal assumptions during the cleaning process.

**Table 1**: Network characteristics of the various matrices.

|  | *JCR 2014* (a) | *Largest component* (b; Figure 1) | *Links ≥ 5* (c; Figure 2) |
| --- | --- | --- | --- |
| **N of journals (nodes)** | 11,149 | 11,143 | 11,087 |
| **Links** | 2,699,210 (10,829 loops removed) | 2,699,210 | 1,196,343 (10,496 loops removed) |
| **Density** | 0.0217 | 0.0217 | 0.0097 |
| **Average (total) degree** | 484.207 | 484.467 | 215.810 |
| **Cluster coefficient** | 0.220 | 0.220 | 0.178 |

Table 1 shows the network characteristics of the various matrices. Column (b)—the largest component of the full set—is similar to column (a) except for the removal of six unconnected journals.[7] Removing the links with values smaller than five can be expected to increase the number of unconnected clusters (see column c in the table).

**4. Decomposition**

*4.1. Which algorithm to use?*

Two routines are available in Pajek for the decomposition: the so-called Louvain algorithm (Blondel *et al*., 2008) and the VOS algorithm. Using 2012 data and a similar design, Leydesdorff & Rafols (2014) found that the Louvain algorithm generated a lower number of singletons than the VOS algorithm, and therefore pursued the analysis with the Louvain algorithm. Table 2 shows the results of two runs using each routine: the numbers of clusters are different between the runs. The quality of the decomposition is measured by the modularity $Q$ when using Blondel *et al*. (2008) and the parameter *VOS Quality* in the other case.

**Table 2:** Decomposition of the largest component of the citation matrix (11,143 journals) using the Louvain or VOS algorithms in Pajek.

| Full matrix | $N$ of clusters | $Q$ or *VOS Quality* |
|---|---|---|
| Blondel *et al*. (2008) | 11 | 0.556 |
|  | 10 | 0.562 |
| VOS (Pajek) | 11 | 0.886 |
|  | 12 | 0.886 |

---

[7] These six journals are: *Edn, Argos-Venezuela, Balt J Econ, Curric Matters, Econtent,* and *Restaurator*.

Although the decompositions are somewhat different, 10 to 12 clusters are found in *all* runs using both algorithms after the single journals are removed from the distributions. These classifications can be compared using chi-square statistics, both in terms of their mutual consistency and in terms of their internal consistency among runs of the same algorithm. Cramer's *V* is a measure of association and is based on the chi-square statistic. Its values for the association range conveniently between zero and one.

The strength of the relationship between the two classifications—of VOSviewer and the Louvain algorithm, respectively—is large: Cramer's $V \approx 0.812$. The internal consistency of the solutions in each of the two routines can be measured by using, for example, five drawings. In the case of the Blondel-algorithm Cramer's $V \approx 0.912$ ($\pm 0.025$ for five drawings) and in the other case $V \approx 0.897$ ($\pm 0.027$). The slightly higher value of Cramer's *V* for the Louvain-algorithm accords with Leydesdorff & Rafols' (2014) preference for this algorithm; but the differences are negligible. However, there remain non-trivial differences in the resulting cluster structures using either algorithm in different runs. The uncertainty thus introduced, is unfortunate from the perspective of the envisaged mapping in layers, since uncertainty will be multiplied at each level of the decomposition.

*4.2. Decomposition with reduced data*

As noted above, we constructed a second matrix in which values of aggregated citations lower than five were considered as noise and therefore removed. Using this matrix, Table 3 provides the analogue of Table 2. Unlike the Blondel algorithm, VOSviewer generates a number of singletons in the decomposition. The numbers minus these singletons are added between brackets.

**Table 3:** Decomposition of the *reduced* citation matrix for 11,087 journals using the Louvain or VOS algorithm in Pajek.

| Matrix with values ≥ 5 (loops removed) | N of clusters | Q or VOS Quality |
|---|---|---|
| Blondel *et al.* (2008) | 11 | 0.581 |
|  | 12 | 0.581 |
| VOS (Pajek) | 31 (17) | 0.923 |
|  | 27 (15) | 0.923 |

The two algorithms thus behave rather differently using the reduced data. When compared with the above analysis of the largest components, Cramer's *V* is lower ($0.68 < V < 0.74$; $p < 0.001$). In the case of comparing two solutions using the same algorithm Cramer's *V* is 0.870 and 0.848, respectively. The lower values indicate that the reliability of the clustering has declined.

Although there are arguments for discarding the tails of the distribution as noise, by doing so one inadvertently introduces a parameter: the relative weights of the tails can be expected to vary among fields of science. In our opinion, one must therefore have strong arguments for introducing this additional parameter; the decomposition did not provide such arguments. On the contrary, the largest component of the full matrix had higher values for the network parameters in Table 1. Do the resulting maps perhaps provide an argument for choosing one of the two algorithms? We focus now on using the VOS algorithm for the decomposition and VOSviewer for the mapping.

## 5. Maps

Using VOSviewer, the maps (see Figures 1 and 2) are based on the largest components of the full matrix and the matrix with reduced data, respectively. In these two cases, VOSviewer distinguishes 11 and 34 clusters, respectively; including 2 and 18 isolates respectively. Since there are two more clusters of only two journals in the reduced case, Figures 1 and 2 show effectively 9 and 14 clusters.

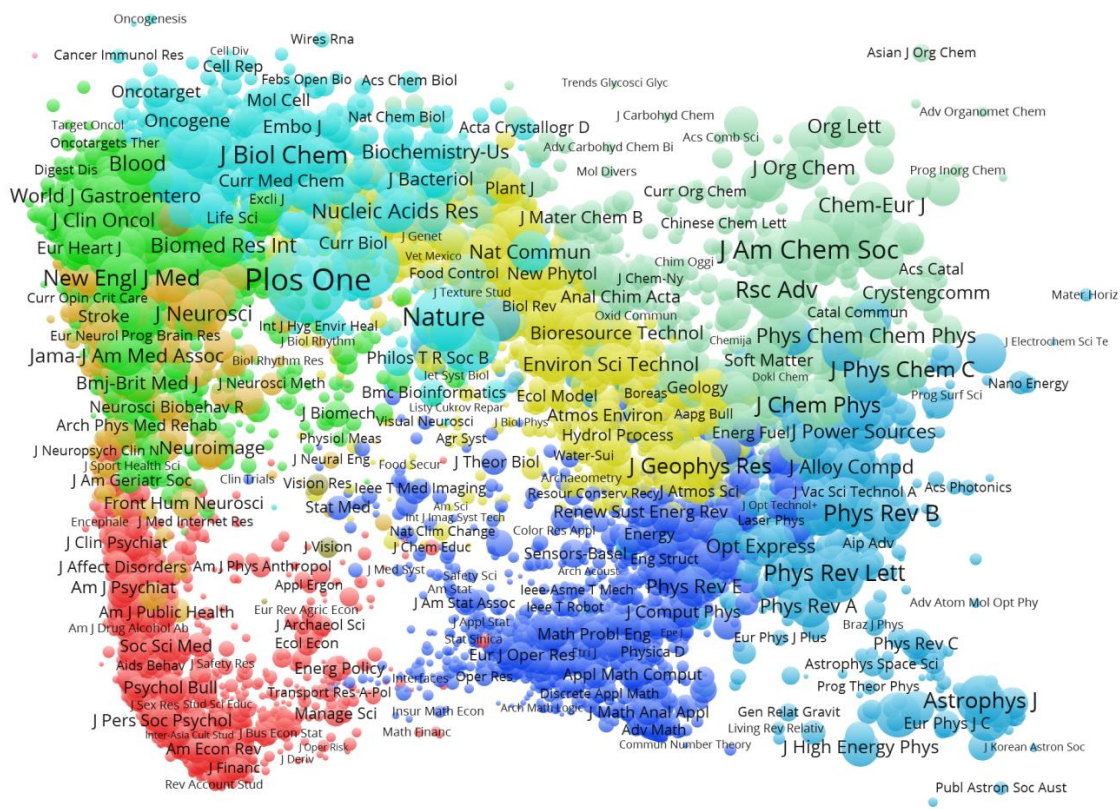

**Figure 1**: Eleven clusters of 11,143 journals (largest component of the JCR matrix); VOSviewer used for classification and visualization. This map can be web-started at

http://www.vosviewer.com/vosviewer.php?map=http://www.leydesdorff.net/journals14/jcr14.txt&cluster_colors=http://www.leydesdorff.net/journals14/colors14.txt&label_size_variation=0.3&zoom_level=1&scale=0.9

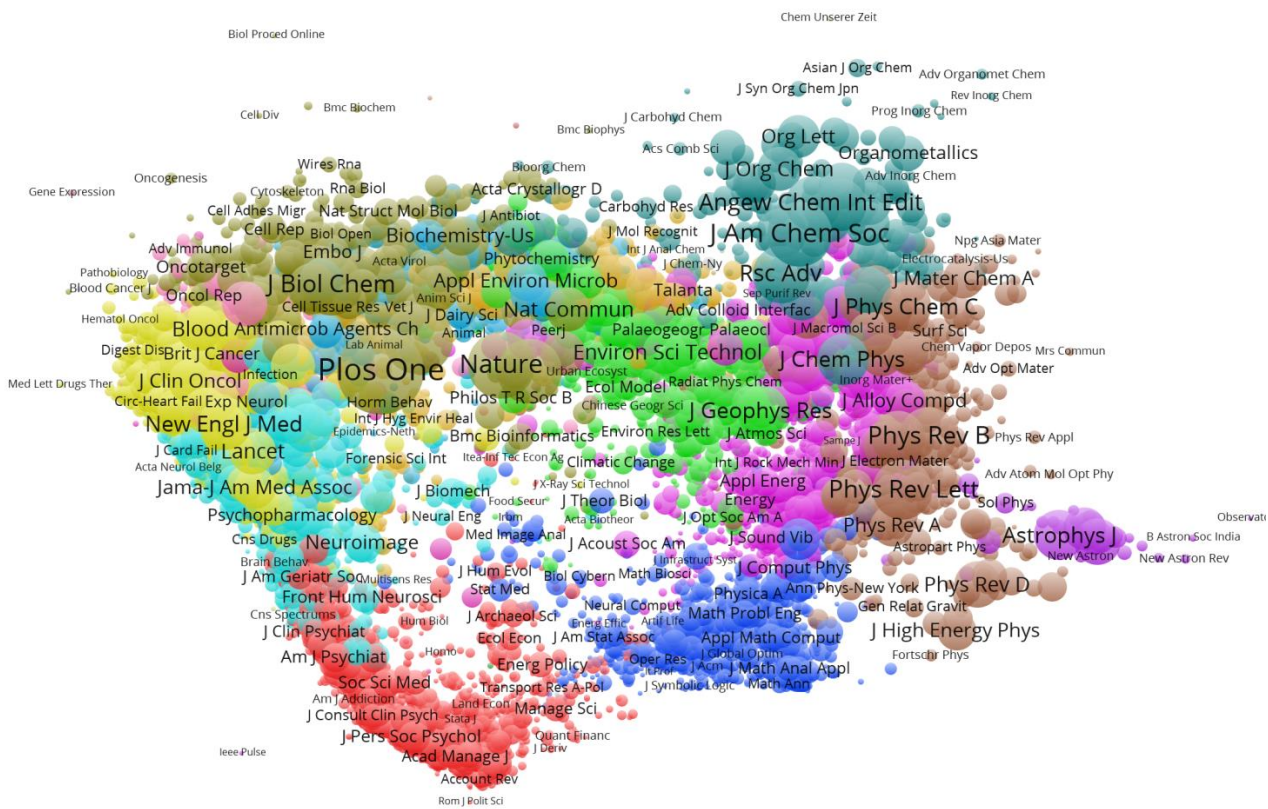

**Figure 2:** Thirty-four clusters of 11,087 journals in the case of reduced data (links of fewer than five were deleted). This map can be web-started at http://www.vosviewer.com/vosviewer.php?map=http://www.leydesdorff.net/journals14/fig2map.txt&label_size_variation=0.3&zoom_level=1&scale=0.9

For example, a group of 54 astrophysics and astronomy journals is distinguished in Figure 2 (at the bottom right), but integrated in the physics group in Figure 1. In both cases, between 56 and 62 journals are set apart as ophthalmology, but in the latter case an additional group of 402 journals is distinguished at the interface between chemistry (notably, analytical chemistry) and the environmental sciences. Otherwise the differences are mainly in the isolates. Comparison of Figures 1 and 2—which are similarly scaled—shows, in our opinion, that Figure 1 is richer: the lobes (e.g., astrophysics journals at the bottom right) are more outreaching. The links below five thus contribute to the quality of the map. Setting a threshold has an adverse effect: without the minor links, which are specific, the larger journals show as more densely packed.[8]

## 6. A pragmatic classification

Using the full matrix, VOSviewer distinguished eleven clusters, among which two are singletons (*Prog Tumor Res* and *Epidemics Neth*; see Table 4). We work with these nine top-layer fields of science. As noted, the designation is not provided by the algorithms, but based on our reading of the algorithmically generated results.

We construct the multi-layered classification by saving the clustering in VOSviewer as a partition file in Pajek (with the extension .clu). The partitioning enables us to extract a subnetwork in Pajek that can be read into VOSviewer after saving it in the .net format.[9] This circle can be reiterated for each next-lower level in the decomposition. Currently, one has to intervene manually for saving the cluster file in the Pajek format. The developers of VOSviewers, however, plan to make it possible to export this information while working from the command line (Nees Jan van Eck, *personal communication*, 3 and 16 May 2016). This may make it possible to automate the production of the classification by using a loop including a macro in Pajek and calling VOSviewer from the command line.

---

[8] VOSviewer symmetrizes the asymmetrical matrix internally by summing the cells ($i,j$) and ($j,i$).
[9] This is one of the options under "Save > Save map" in VOSviewer. This partition information can be read into Pajek and then be used for extraction of each of the clusters from the network using "Operations > Network + Partition > Extract Subnetwork."

*6.1.   Top-layer distinction among nine fields of science*

As noted, nine fields are distinguished in the initial decomposition of the largest component of the grand matrix (*N* of journals is 11,141; see Table 4). The group of social-science journals is by far the largest grouping. The journals in the social sciences obviously share a citation pattern that is different from the other groups. Ophtalmology includes a relatively small set of journals. In Figure 1, this cluster is difficult to track without first zooming in on the brown-colored journals at the interface between the bio-medical journals (light blue) and medical journals (green) in the upper left quadrant.

**Table 4**: Nine fields of science distinguished in JCR at the top level

| Field | N |
|---|---:|
| Social Sciences | 3,131 |
| Medicine | 1,943 |
| Computer Science | 1,939 |
| Environmental | 1,911 |
| Chemistry | 684 |
| Bio-Medical | 672 |
| Physics | 462 |
| Neuro Sciences | 343 |
| Ophthalmology | 56 |
| Sum | 11,141 |

As noted above, we pursue the analysis using *JASIST* and *Scientometrics* as our leads for the decomposition. The decomposition of the other branches is equally possible, as we will demonstrate using other, more qualitatively oriented journals in science and technology studies as an example (Leydesdorff & Van den Besselaar, 1997; Wyatt *et al*., 2016). Table 5 provides a summary of the decompositions that will be pursued. We envisage completing the classification in a next project.

**Table 5:** Decomposition of the JCR at different levels (fields, subfields, specialties)

| Fields  Subfields/Disciplines  Specialties | N of journals |
|---|---|
| 1. Social Sciences | 3,131 |
| 2. Medicine | 1,943 |
| 3. Computer Science | 1,939 |
| 4. Environmental Sciences | 1,911 |
| 5. Chemistry | 684 |
| 6. Bio-Medical Sciences | 672 |
| 7. Physics | 462 |
| 8. Neuro Sciences | 343 |
| 9. Ophthalmology | 56 |
| Decomposition of 1. Social Sciences | |
|    1.1. Discipline-oriented social sciences | 1,008 |
|    1.2. Application-oriented social sciences | 385 |
|    1.3. Health | 345 |
|    1.4. Economics | 335 |
|    1.5. Mental Health | 267 |
|    1.6. Administration | 255 |
|    1.7. Language | 188 |
|    1.8. Psychology | 146 |
|    1.9. Law | 117 |
|    1.10. Library & Information Science | 52 |
|    1.11. Transport | 33 |
| Decomposition of 1.1. Discipline-oriented social sciences | |
|      1.1.1. Anthropology | 258 |
|      1.1.2. Sociology | 143 |
|      1.1.3. History and Philosophy of Science | 128 |
|      1.1.4. Geography | 101 |
|      1.1.5. International Relations | 100 |
|      1.1.6. Political Science | 78 |
|      1.1.7. Environmental | 69 |
|      1.1.8. International Law | 63 |
|      1.1.9. Communication Studies | 43 |
|      1.1.10. Archaeology | 25 |
| (…) | |
| Decomposition of 1.1.3. History and Philosophy of Science | |
|        1.1.3.1. Science Studies (STS) | 20 |
|        1.1.3.2. Science Education | 10 |
|        1.1.3.3. History of Science | 35 |
|        1.1.3.4. Health Ethics | 26 |
|        1.1.3.5. Socio-biology | 2 |
|        1.1.3.6. Philosophy of Science | 18 |
|        1.1.3.7. Ethics and Social Philosophy | 17 |
| (…) | 28 |
| Decomposition of 1.10. Library & Information Science | 9 |
|      1.10.1. Library Science | 5 |
|      1.10.2. Information & Organization | 3 |
|      1.10.3. Publishing | 3 |
|      1.10.4. *ASLIB* journals | 3 |
|      1.10.5. Scientometrics | 1 |
|      1.10.6. *JACS + Z Bibl Bibl* | |
|      1.10.7. *Can J Inform Lib Sci* | |

*6.2.    Further decomposition of the set of 3,131 social-science journals*

We pursued the decomposition using the choices and procedures specified above. Figure 3 shows a map of the eleven clusters of social-science journals that are summarized in Table 6. The first and largest set is composed of disciplinarily oriented journals in the social sciences ($N = 1,008$) with a citation pattern different from some other disciplinary clusters (e.g., economics and psychology) and some fields of application (e.g., "health" and "transport"). "Library & information science" is distinguished at this level as a group of 52 journals which we will further analyze in the next section.

**Table 6:** Decomposition of the set of 3,131 journals in the social sciences

| Subfields | N |
|---|---:|
| Discipline-oriented social science | 1,008 |
| Application-oriented social science | 385 |
| Health | 345 |
| Economics | 335 |
| Mental Health | 267 |
| Administration | 255 |
| Language | 188 |
| Psychology | 146 |
| Law | 117 |
| Library & Information Science | 52 |
| Transport | 33 |
| Sum | 3,131 |

Table 6 shows that clusters can sometimes be designated as disciplines (e.g., economics, psychology, law), but in other cases as fields of application (e.g., transport, health). As noted, the designation is not a result of the analysis, but based on the semantics which we as analysts use for understanding the algorithmic results; in other contexts, one may wish to use other terminology.

**Figure 3**: Eleven clusters of citation patterns among 3,131 journals in the social sciences. This figure can be web-started at http://www.vosviewer.com/vosviewer.php?map=http://www.leydesdorff.net/journals14/level2/sosci.txt&label_size_variation=0.4&scale=0.9

*6.3. Decomposition and map of 52 journals in library and information science*

The 52 journals in library and information science contain a largest cluster of 28 journals which can be denoted as "library science" *sensu stricto*. Among the other 24 journals, three are identified as a separate group which we denote as "bibliometrics." These are *Scientometrics, Journal of Informetrics,* and *Research Evaluation. JASIST*—represented both as the *Journal of the American Society for Information Science and Technology* (that is the name until 2014) and the *Journal of the Association for Information Science and Technology* (that was the name since 2014)—forms a separate group with *Z Bibl Bibl.* The *Malays J Lib Inf Sci* is placed in a cluster with the two ASLIB journals in the database: *ASLIB J Inform Manag* and *ASLIB Proc.* The *Can J Inform Lib Sci* is a singleton.

In Appendix 1, these 52 journals are compared with the 85 journals subsumed under the category "information science & library science" in WoS.[10] *Issues Sci Technol* and *J Legal Educ* are not counted as LIS in WoS, but belong to the specialty in terms of their cited/citing patterns in our classification. However, 33 journals in the WoS category are not counted as LIS using our map (Figure 4) and classification. These journals are mainly about the management of information systems, such as *MIS Quarterly.* Subsuming these two groups of journals into a single WC on the basis of the word "information" has been a major problem in the WoS classification (Leydesdorff & Bornmann, 2016). The two groups are very different in terms of citation behavior. This entire group is classified differently in this decomposition: under the category 1.6 in Table 5, which is labeled "Administration" and contains 255 journals in total.

---

[10] In WoS, this category is abbreviated as "NU".

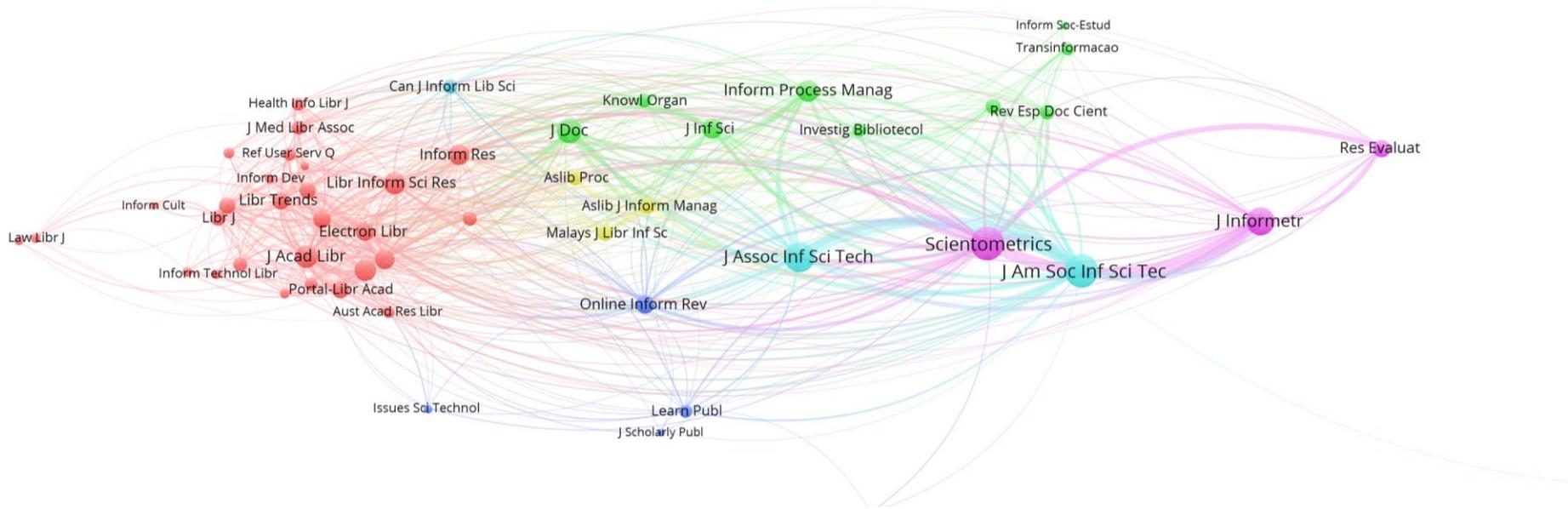

**Figure 4**: Map of 52 journals classified as Library and Information Science. This file can be web-started at http://www.vosviewer.com/vosviewer.php?map=http://www.leydesdorff.net/journals14/level3/lis52map.txt&network=http://www.leydesdorff.net/journals14/level3/lis52net.txt&zoom_level=1.5&label_size_variation=0.3&scale=1.1&colored_lines&curved_lines&n_lines=10000

*6.4.   The disciplinary-organized group of the social sciences (cluster 1.1)*

We labeled the largest group at the second level as "discipline-oriented social sciences" ($N = 1,008$). Note that the disciplines of economics (335 journals) and psychology (146 journals) are already separated out at this level, as was the group of 52 LIS journals discussed above. The next decomposition of the largest group at the second level provides a structure of seven disciplines in the social sciences and three in the humanities. These distinctions are in our opinion very meaningful. Figure 5 provides the map and Table 7 the categories and numbers of journals involved.

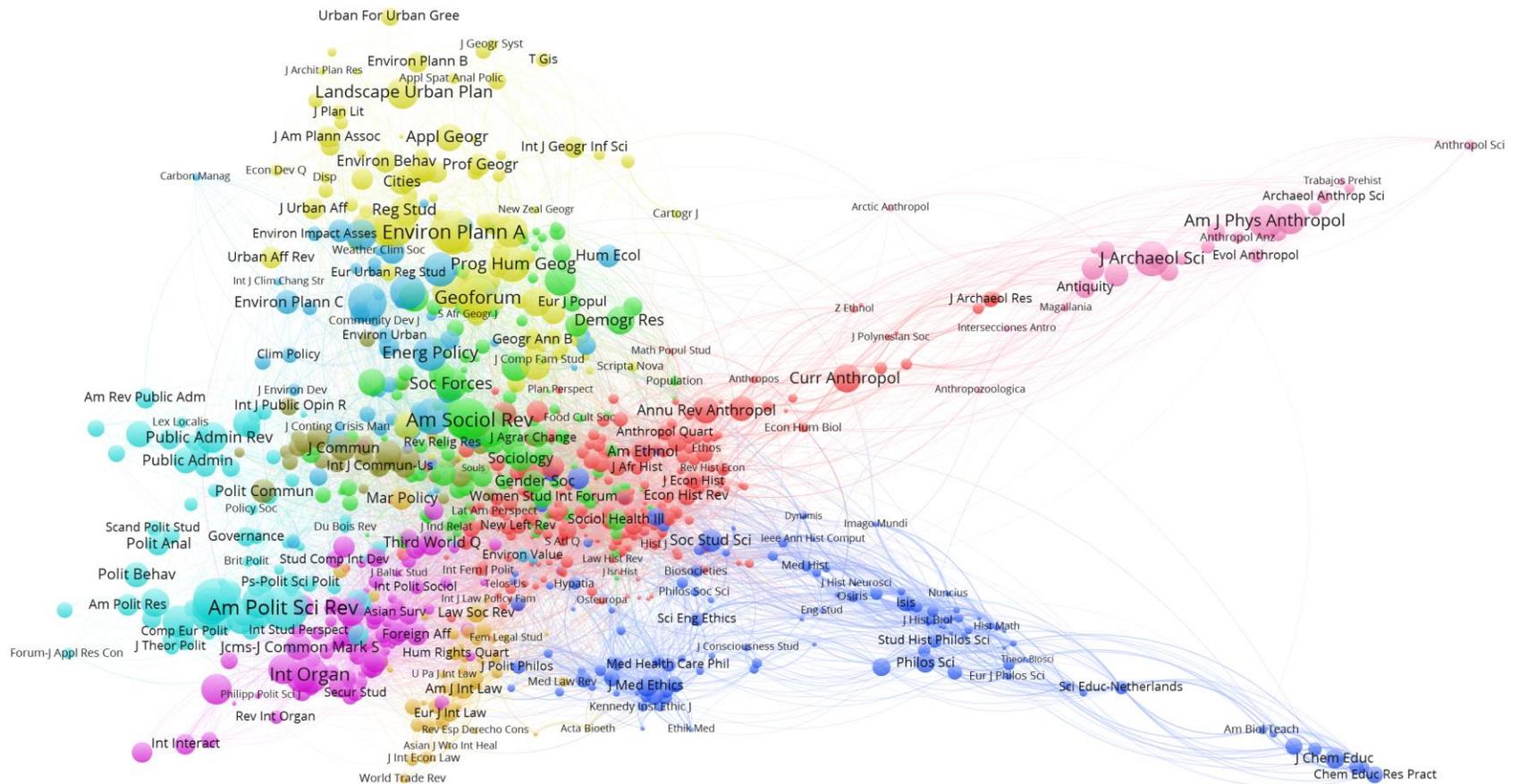

**Figure 5**: Ten clusters among 1,008 journals in disciplinarily organized social and cultural sciences. This file can be web-started at http://www.vosviewer.com/vosviewer.php?map=http://www.leydesdorff.net/journals14/level3/sosci1_map.txt&label_size_variation=0.4&scale=0.9&colored_lines&n_lines=10000&normalized_lines&curved_lines

**Table 7**: Decomposition of the discipline-oriented group of 1,008 journals in the social sciences

| Disciplines | N |
|---|---|
| Anthropology | 258 |
| Sociology | 143 |
| History and Philosophy of Science | 128 |
| Geography | 101 |
| International Relations | 100 |
| Political Science | 78 |
| Environmental | 69 |
| International Law | 63 |
| Communication Studies | 43 |
| Archaeology | 25 |
|  | 1,008 |

The three groups of journals in the humanities make the map excentric. Most pronouncedly the archaeology group ($N = 25$) at the top right is hardly connected to other groups except anthropology. At the bottom right, one observes a large group of journals involved in the study of science and technology from different perspectives (history, philosophy, education, etc.). The law journals ($N = 63$) shape a lobe at the bottom of the figure. We pursue the decomposition of the history and philosophy of science (HOPOS) group in order to show the position and fine-structure of science and technology studies (STS). We return to the distinction between "international law" ($n = 63$) at this level and the previously distinguished group of "law" journals ($n = 117$) in the section below (section 7) about limitations.

## 6.5. *History and Philosophy of Science (HoPoS) and Science Studies (STS) (decomposition of cluster 1.1.3 at level 4)*

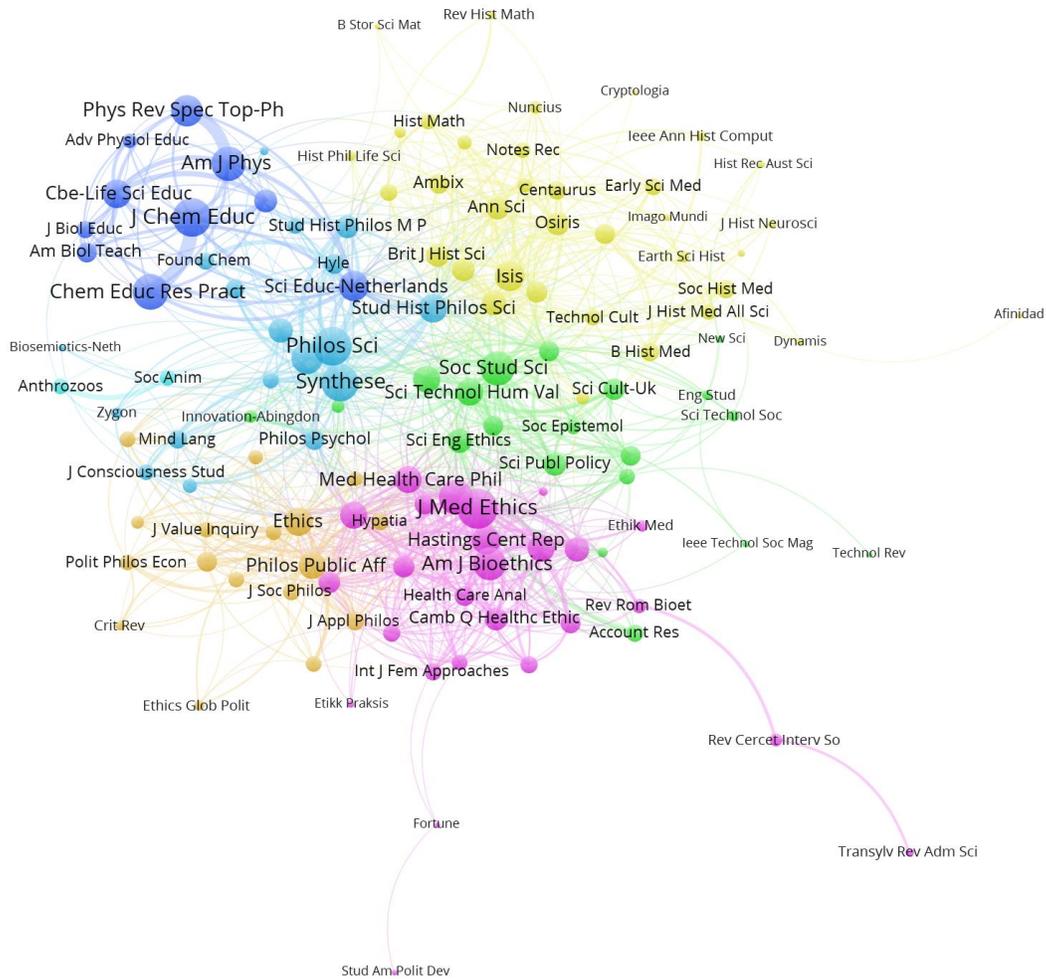

**Figure 6**: Eight clusters of 128 journals in *history and philosophy of science (HoPoS)*. Layout according to Kamada & Kawai (1989), clustering according to Blondel *et al.* (2008), using Pajek. The map can be web-started at
http://www.vosviewer.com/vosviewer.php?map=http://www.leydesdorff.net/journals14/level4/sts_map.txt&network=http://www.leydesdorff.net/journals14/level4/sts_net.txt&label_size_variation=0.4&scale=0.9&cluster_colors=http://www.leydesdorff.net/journals14/level4/sts_col.txt&colored_lines&n_lines=10000&curved_lines

**Table 8**: Decomposition of the group of 128 journals in history and philosophy of science (HoPoS)

| Specialty | N | WC |
|---|---|---|
| Science Studies (STS) | 20 | 9 |
| Science Education | 10 | 1 |
| History of Science | 35 | 34 |
| Health Ethics | 26 | 1 |
| Socio-Biology | 2 | 0 |
| Philosophy of Science | 18 | 11 |
| Ethics and Social Philosophy | 17 | 1 |
|  | 128 | 57 |

In the case of Figure 6, we used another algorithm for the layout in Pajek (Kamada & Kawai, 1989) because the mapping of VOSviewer was less informative.[11] Note the ease of using different algorithms whenever convenient.[12]

This group of 128 journals can be compared with the category "History & Philosophy of Science" in WoS containing 67 journals, of which 57 are included among these 128. The additional column in Table 8 teaches us that the health ethics and the science education journals in particular are located differently according to the WoS classification.

*6.6.    Science Studies (STS) (level 5; 20 journals)*

Let us pursue the analysis in this case also at the next-lower level of the 20 journals labeled above as STS. The distinctions are now fine-grained and precise. The group on the right is focused on ethical discussions about science-and-society issues in engineering and engineering education. Radical STS is concentrated in a group of five journals around *Social Studies of Science* (green). *Science & Public Policy, Minerva,* and *Public Understanding of Science* form the core of a third group (blue) that is further extended with two minor journals. Discussions at the philosophical level are indicated as two journals (*Soc Epistemol* and *Sci Technol Soc*) represented by pink-colored nodes (Figure 7).

---

[11] After web-starting Figure 6, one can obtain a very informative map of this domain by clicking the tab "Analysis"; uncheck "Use default options;" change "Repulsion" to zero; and "Update Layout" (Ludo Waltman, *personal communication*, 9 July 2016).
[12] In this case, one draws the figure first within Pajek and then exports to VOSviewer using "Export > 2D > VOSviewer" from the drawing screen in Pajek.

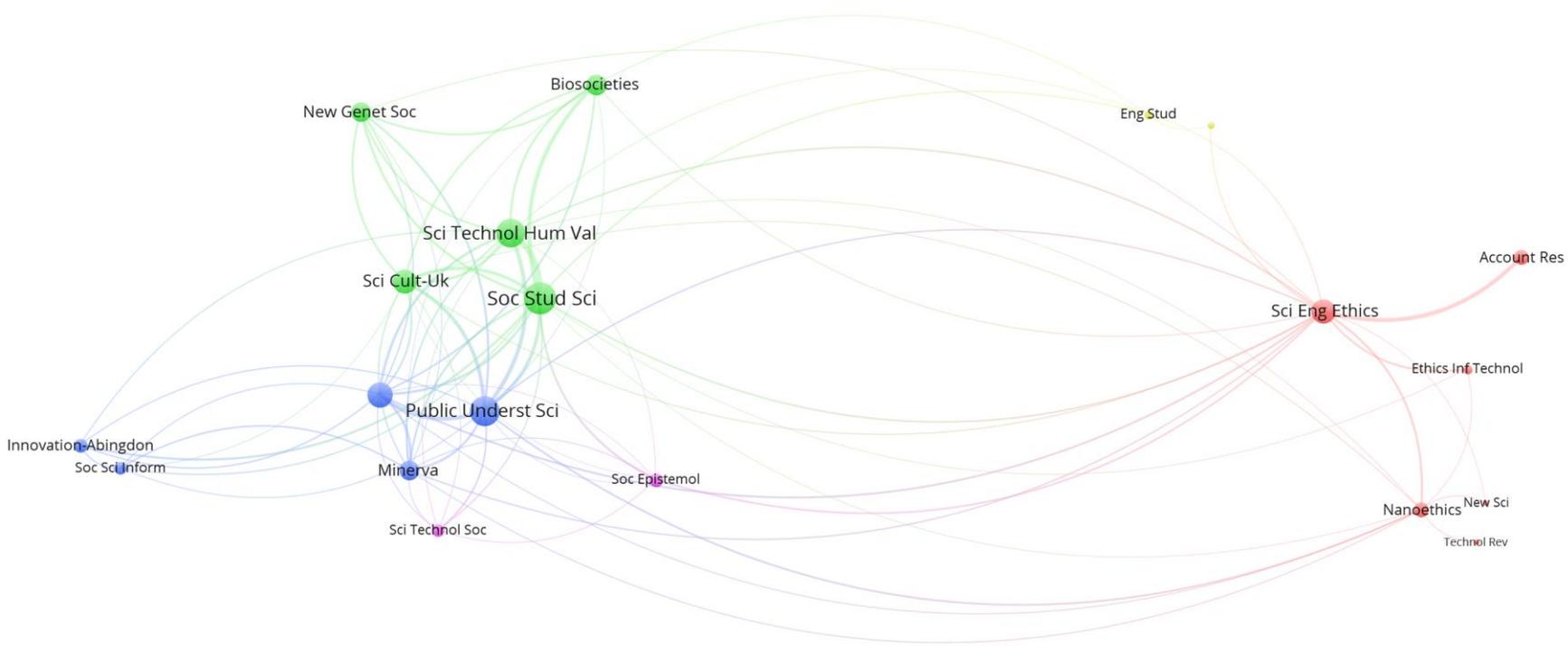

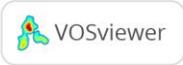

**Figure 7**: Twenty journals in the specialty of STS/sociology of science; four clusters distinguished. This figure can be web-started at http://www.vosviewer.com/vosviewer.php?map=http://www.leydesdorff.net/journals14/level5/sss_map.txt&network=http://www.leydesdorff.net/journals14/level5/sss_net.txt&label_size_variation=0.4&scale=1.1&colored_lines&n_lines=10000&curved_lines

## 7. Limitations

Before drawing conclusions and summarizing, let us turn to some limitations of the empirical analyses in more detail. The main problem with hierarchical and divisive clustering is that each journal has to be uniquely attributed to a single class. In the case of multidisciplinary journals, attribution to more than one category may be desirable. Secondly, the classes impose a structure with divisions among journals which in other dimensions may be more akin. We elaborate on (*i*) the problem of multi-disciplinary journals by focusing on *PLoS ONE* as the journal which is programmatically not bound to a single discipline and (*ii*) the problem of perhaps disturbing divides by studying the relations between the two classes of law journals distinguished in Table 5 (Classes 1.9 and 1.1.8).

### 7.1. *PLoS ONE*

Table 9 shows the decomposition of the cluster containing among other journals *PLoS ONE*. On the basis of the prevailing pattern in its citation, *PLoS ONE* is categorized as a molecular-biology journal and positioned very close to *Nature* and *Science* in Figure 8. This group of 80 journals can be considered as a reference set, in our opinion. The further decomposition leads to the placement of *PLoS ONE* in a group of eight molecular genetics journals. Figure 8 shows the structure of the 80 journals in Class 6.3.

**Table 9**: The classification of *PLoS ONE* at different levels.

| | |
|---|---|
| 6. Bio-Medical Sciences | 672 |
| 6.3. Molecular Biology | 80 |
| 6.3.5 Molecular Genetics | 8 |
| 6.3.5.1 Genomics | 5 |

Using this classification, one can thus determine relevant reference sets of journals for *PLoS ONE* in terms of its pattern of citation relations at different levels of granularity. In our opinion, the group of eighty journals will be the better choice in most evaluations—for example, to determine the top-10% most-highly cited papers in a journal-based reference set—but this choice firstly depends on the research question. The classification only clarifies the options.

**Figure 8**: Eighty interdisciplinary and bio-medical journals co-classified with *PLoS ONE* at the second level. This figure can be web-started at
http://www.vosviewer.com/vosviewer.php?map=http://www.leydesdorff.net/journals14/plosone/plosmap.txt&network=http://www.leydesdorff.net/journals14/plosone/plosnet.txt&label_size_variation=0.4&scale=1.1&colored_lines&n_lines=10000&curved_lines&zoom_level=2

*7.2. The classification of law journals*

In Table 5, two groups of law journals were differently classified: one group of 117 journals was classified at the second level as "Law" (Class 1.9) and a second group of 63 journals as "International Law" in Class 1.1.8 (as one of the discipline-oriented social sciences). We noted above that the latter group shapes a lobe at the bottom of Figure 5. One can raise the question of how journals in these two classes relate.

By combining the two partitions 1.9 and 1.1.8, one can extract this combined set of (117 + 63 =) 180 journals from the matrix.[13] In Figure 9, three main clusters are distinguished: one on the left side of 51 journals of which 31 have the words "Law Review" in the title; a second one (on the right side) of 59 journals in criminology; and a third one which virtually coincides with Class 1.1.8 designated above as "international law".[14] This latter group includes journals about European law systems and human right issues, whereas the journals with "law review" in their titles are mainly American. The wider scope of law as a system of legislation and governance makes the international group closer to the social sciences in terms of aggregated citation relations than to the more specialized law journals in the other two groups.

---

[13] In Pajek, one selects to this end "Operations > Network + Partition > …".
[14] Two journals of both classes 1.9 and 1.1.8 are organized in a fourth group (indicated with yellow nodes) of ethical and legal studies.

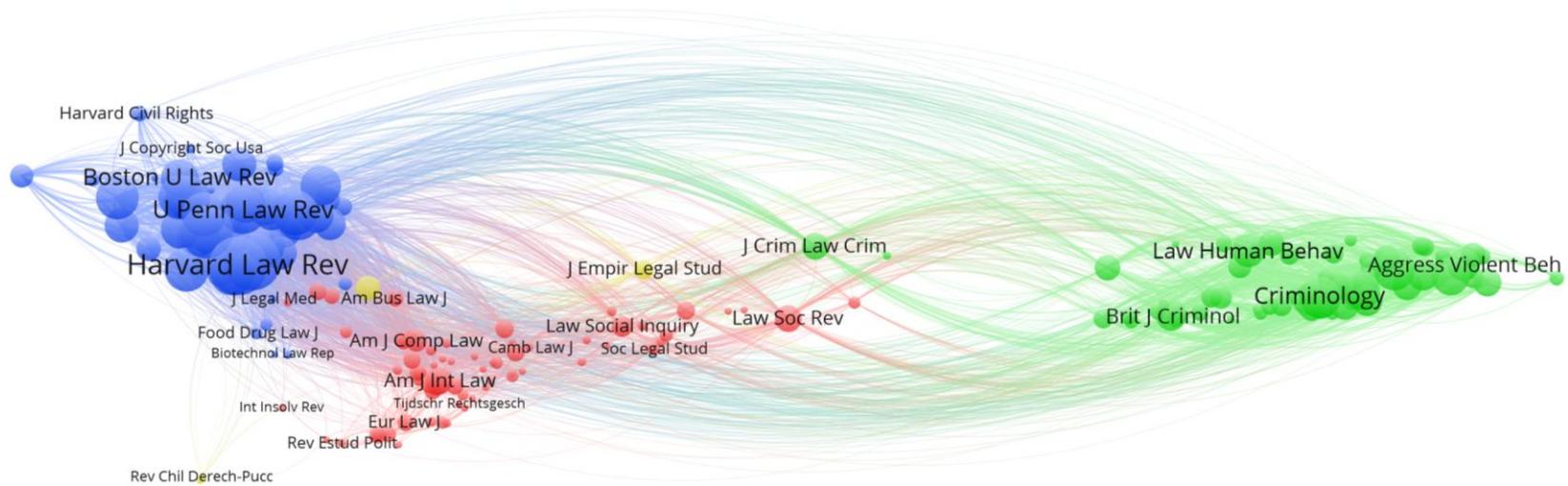

**Figure 9**: Journals in law (*n* = 117; Class 1.9) and international law (*n* = 63; Class 1.1.8) combined. The figure can be web-started at http://www.vosviewer.com/vosviewer.php?map=http://www.leydesdorff.net/journals14/law/law_180_map.txt&network=http://www.leydesdorff.net/journals14/law/law_180_net.txt&label_size_variation=0.45&scale=1.25&colored_lines&n_lines=10000&curved_lines&zoom_level=1

The partition of 180 journals can be exported in Pajek to an SPSS syntax file.[15] After reading into SPSS, one can, for example, factor analyze the citation matrix. A four-factor solution (explaining 50.8% of the variance), for example, teaches us that the group of journals with "law review" in the title loads on factor 1; criminology journals on factor 2; and "international law" journals—as defined above—on factor 3. Factor 4 extracts forensic journals as a separate group. In a three-factor solution, this latter group would be loading positively on factor 2 (criminology), but negatively on factor 3 (international law). Thus, one can specify the factorial complexity of the relations between groups that were divided by the decomposition algorithm.

Different from factor analysis, the decomposition algorithms can process large networks virtually without systems limitations. One can thus generate partitions that can be analyzed in greater detail using, for example, factor analysis. However, the number of factors to be extracted has to be set among other parameters by the analyst, whereas the decomposition algorithms guide us in a meaningful breakdown of the agglomerate.

### 7.3. *The dynamic extension of the classification*

One can consider the stability of the WCs over time as one of their advantages in the practice of evaluative bibliometrics. This stability is a consequence of the deliberate choice of the database producer: during several decades, the WCs were incrementally improved and extended (Bensman &Leydesdorff, 2009). In 2005, for example, a category for nano-science and nano-technology was added. The here proposed classification, however, enables us to use the current standards and understanding as references different from a historical understanding, and to backtrack from our present definitions. Thus, we inverse the arrow of time and can signal from a perspective of hindsight when new developments have become important in terms of the database (Leydesdorff, 2002).

---

[15] In Pajek, one can use: Tools > SPSS > Send to SPSS.

One may wish to have a dynamic classification which develops with the database. The pragmatic approach chosen in this paper does not provide such a classification. Using other data (e.g., other years), one can expect globally similar, but in details potentially different results. In a first exploration, for example, backward extension to 2013 data taught us that in this year a cluster of 43 astronomy and astrophysics journals is distinguished from the physics journals in the first iteration, whereas these two groups were a single top-level group in 2014 ("Physics"). The classification is sensitive to detailed (changes in) citation patterns as we saw above for the case of "international law" versus "law" journals. In other years, the distinctions may be different— for example, because of special issues of journals—and a classification for this other year would also be different. One can only compare across years given a classification.[16]

## 8. Conclusions and discussion

Field-classification systems are used in bibliometrics for normalizing citation counts. The best known and most frequently used system is the WCs. Using VOSviewer and Pajek, this study examines options for developing a new classification system of journals on the basis of the aggregated journal-journal citation data provided in the two JCRs.[17] The social sciences formed the largest group in the first round: more than 3,000 of the 11,000+ journals exhibit a specific citation pattern different from the other sciences. At a next round, more than 1,000 of these 3,000+ journals form the core journals of the various disciplines in the social sciences; the others are application oriented. One of the theory-oriented groups was further analyzed in this study with a focus on science and technology studies. Note how differently journals like *Scientometrics* or *Social Studies of Science* are positioned in this classification system despite their common background in science studies (Leydesdorff & Van den Besselaar, 1997; Wyatt *et al*., 2016).

---

[16] The network analysis and visualization program *visone* contains a routine for dynamic multi-dimensional scaling. This routine minimizes stress both over time and at each moment of time, but it does not calculate a clustering after each time step. Furthermore, the capacity in terms of the numbers of nodes and links is limited (Brandes *et al.*, 2012; Leydesdorff & Schank, 2008).

The proposed classification is one among other possible ones (e.g. systems which classify articles algorithmically on the basis of direct citations). In this early stage of development, the proposed classification offers also a research and analysis tool:

1. Given the citation matrix, the generation of the hierarchical dendrogram can virtually be automated; the procedures can be used for other matrices such as citation data for other years or other databases (e.g., Scopus);
2. The matrix of aggregated citations among journals is "nearly decomposable" (Simon, 1962; 1973): in addition to strongly interrelated clusters of journals, some journals span across these horizontal differentiations, for example, as structures of elite journals or, in other words, vertical differentiations (Leydesdorff, 2006). Any hierarchical classification obviously reduces this complexity and remains one among other possible classifications. Yet, the classification proposed here is not arbitrary or analyst-dependent, since solely based on an algorithmic analysis of the underlying citation distributions (cf. Rafols and Leydesdorff, 2009);

---

[17] A JCR of the Arts & Humanities Citation Index is not available; see Leydesdorff, Hammarfeldt & Salah (2011) for a journal-mapping of this index.

3. The labeling is free and left to the analyst; the analyst is both challenged and legitimated to choose an appropriate designation given his/her research questions and objectives;

4. The attribution of journals to a single categorization provides a heuristics; the user may consciously wish to deviate from the algorithmic results and thus generate a specifiable "indexer effect";

5. Classes and subclasses can be combined and exported to SPSS or R for further statistical analysis;

6. The algorithmic results can be reproduced in other contexts since the problem of the randomness in each run is circumvented.

As these points reveal, the results on the basis of the aggregated journal-journal citation data can be considered as providing a base line for more precise and informed classification. No subjective elements are introduced *ex ante* and the problem of randomness in the initial seed that hitherto generated uncertainty in the results and made them irreproducible from run to run—as shown above when comparing the Blondel algorithm with VOSviewer for the decomposition—has been stabilized. The results are both visually and statistically challenging.

We elaborated the proposed system in one branch given our interest in LIS and STS; but there are no reasons why this could not be done for the other eight branches which were first distinguished as main fields. At the second level, the designation in terms of subfields and disciplines is more complex: "economics" and "psychology," for example, are distinguished at the same (second) level as fields of application like "health" and "transport." Other disciplines (e.g., "sociology" or "anthropology") are distinguished only at the third level. In summary, the database contains a set of organized densities of citations. Words such as "subfields" and "disciplines" can be considered as part of the semantics that we as analysts bring to the data.

The resulting clustering provides a stable representation of the journal structure in the database. All data is exploited; apart from the parameters built into VOSviewer—we used default values—no further decisions implying parameters are made. We thus made an attempt to solve the problem formulated by Rafols & Leydesdorff (2009) that none of the content-based or algorithmically generated classifications were sufficiently precise (cf. Thijs *et al*., 2013). The content-based ones suffer from indexer effects and the algorithmically generated ones were vulnerable to random factors. When maps and classifications are uncertain, reliable normalizations of scientometric indicators are impossible because different reference sets are possible for the same set of documents under study. The top-10% of most highly cited papers, for example, can be different given slightly different reference sets.

The proposed solution is based on commonalities in citation behavior, but remains a hierarchical and divisive clustering tree. Journals are assigned to a single category each, but journals themselves are not homogenous units of analysis (Klavans & Boyack, 2015). The clustering is based on the main trends in the citation distribution after aggregation of the distributions at the level of articles. However, an individual article may differ substantially in its citation behavior or being-cited characteristics from the main trend in the journal in which it is published. We demonstrated the problem by analyzing the multi-disciplinary journal *PLoS ONE*. Thus, our results confirm the results of other studies which questioned the use and accuracy of journal classification schemes and their usefulness for evaluation because they may provide far less accurate representations of knowledge than document-level classifications (Waltman & Van Eck, 2012; Boyack, & Klavans, 2010; Boyack et al., 2011).

This classification is not in terms of cognitive content, but in terms of common patterns in citation behavior; it can therefore serve for the purpose of normalization in bibliometric evaluations. It should be investigated in future studies, whether the proposed classification leads to more reliable, fair and valid normalization results and how the problem of multi-disciplinary journals can be handled. The substantive interpretation of the proposed classifications by the analyst—the labeling—however, is not directly relevant to the bibliometric results. This caveat has a normative implication: one would like to use (change in) the journal map as a baseline for the evaluation of policy initiatives (Leydesdorff, 1986; Studer & Chubin, 1980, pp. 269 ff.). Policy initiatives, however, are based on considerations other than citation behavior. We would therefore expect these maps to be of limited value for this purpose. The overall map (Figure 1), however, provides an excellent platform for portfolio analysis (Leydesdorff, Heimeriks, & Rotolo, 2016). Using the distances on the map, one can also elaborate the ecological disparity and thus compute, for example, Rao-Stirling diversity (Rafols & Meyer, 2010; Stirling, 2007; Zhang *et al.*, 2006).

**Acknowledgement**

We thank Thomson-Reuters for providing us with the JCR data.

**Appendix 1**: Comparison of the LIS category (52 journals) with the WC "information science & library science" (85 journals).

| VOSviewer | WoS |
|---|---|
| Afr J Libr Arch Info | Afr J Libr Arch Info |
| Aslib J Inform Manag | Aslib J Inform Manag |
| Aslib Proc | Aslib Proc |
| Aust Acad Res Libr | Aust Acad Res Libr |
| Aust Libr J | Aust Libr J |
| Can J Inform Lib Sci | Can J Inform Lib Sci |
| Coll Res Libr | Coll Res Libr |
|  | Data Base Adv Inf Sy |
|  | Econtent |
| Electron Libr | Electron Libr |
|  | Ethics Inf Technol |
|  | Eur J Inform Syst |
|  | Gov Inform Q |
| Health Info Libr J | Health Info Libr J |
| Inf Tarsad | Inf Tarsad |
| Inform Cult | Inform Cult |
| Inform Dev | Inform Dev |
|  | Inform Manage-Amster |
|  | Inform Organ-Uk |
| Inform Process Manag | Inform Process Manag |
| Inform Res | Inform Res |
|  | Inform Soc |
| Inform Soc-Estud | Inform Soc-Estud |
|  | Inform Syst J |
|  | Inform Syst Res |
|  | Inform Technol Dev |
| Inform Technol Libr | Inform Technol Libr |
|  | Inform Technol Peopl |
|  | Int J Comp-Supp Coll |
|  | Int J Geogr Inf Sci |
|  | Int J Inform Manage |
| Investig Bibliotecol | Investig Bibliotecol |
| Issues Sci Technol |  |
| J Acad Libr | J Acad Libr |
|  | J Am Med Inform Assn |
| J Am Soc Inf Sci Tec | J Am Soc Inf Sci Tec |
| J Assoc Inf Sci Tech | J Assoc Inf Sci Tech |
|  | J Assoc Inf Syst |
|  | J Comput-Mediat Comm |
| J Doc | J Doc |
|  | J Glob Inf Manag |
|  | J Glob Inf Tech Man |
|  | J Health Commun |
| J Inf Sci | J Inf Sci |
|  | J Inf Technol |
| J Informetr | J Informetr |
| J Legal Educ |  |
|  | J Knowl Manag |
| J Libr Inf Sci | J Libr Inf Sci |
|  | J Manage Inform Syst |
| J Med Libr Assoc | J Med Libr Assoc |
|  | J Organ End User Com |
| J Scholarly Publ | J Scholarly Publ |
|  | J Strategic Inf Syst |
|  | Knowl Man Res Pract |
| Knowl Organ | Knowl Organ |
| Law Libr J | Law Libr J |
| Learn Publ | Learn Publ |
| Libr Collect Acquis | Libr Collect Acquis |
| Libr Hi Tech | Libr Hi Tech |
| Libr Inform Sc | Libr Inform Sc |
| Libr Inform Sci Res | Libr Inform Sci Res |
| Libr J | Libr J |
| Libr Quart | Libr Quart |
| Libr Resour Tech Ser | Libr Resour Tech Ser |
| Libr Trends | Libr Trends |
| Libri | Libri |
| Malays J Libr Inf Sc | Malays J Libr Inf Sc |
|  | Mis Q Exec |
|  | Mis Quart |
| Online Inform Rev | Online Inform Rev |
| Portal-Libr Acad | Portal-Libr Acad |
| Prof Inform | Prof Inform |
| Program-Electron Lib | Program-Electron Lib |
| Ref User Serv Q | Ref User Serv Q |
| Res Evaluat | Res Evaluat |
|  | Restaurator |
| Rev Esp Doc Cient | Rev Esp Doc Cient |
|  | Scientist |
| Scientometrics | Scientometrics |
| Serials Rev | Serials Rev |
|  | Soc Sci Comput Rev |
|  | Soc Sci Inform |
|  | Telecommun Policy |
|  | Telemat Inform |

| Transinformacao | Transinformacao |
| Z Bibl Bibl | Z Bibl Bibl |